\documentstyle[12pt]{article}

\catcode`\@=11
\long\def\@makefntext#1{
\protect\noindent \hbox to 3.2pt {\hskip-.9pt
$^{{\ninerm\@thefnmark}}$\hfil}#1\hfill}		

\def\@makefnmark{\hbox to 0pt{$^{\@thefnmark}$\hss}}  

\def\ps@myheadings{\let\@mkboth\@gobbletwo
\def\@oddhead{\hbox{}
\rightmark\hfil\ninerm\thepage}
\def\@oddfoot{}\def\@evenhead{\ninerm\thepage\hfil
\leftmark\hbox{}}\def\@evenfoot{}
\def\sectionmark##1{}\def\subsectionmark##1{}}

\renewcommand{\thefootnote}{\fnsymbol{footnote}}

\newcounter{sectionc}\newcounter{subsectionc}\newcounter{subsubsectionc}
\renewcommand{\section}[1] {\vspace*{0.6cm}\addtocounter{sectionc}{1}
\setcounter{subsectionc}{0}\setcounter{subsubsectionc}{0}\noindent
	{\normalsize\bf\thesectionc. #1}\par\vspace*{0.4cm}}
\renewcommand{\subsection}[1] {\vspace*{0.6cm}\addtocounter{subsectionc}{1}
	\setcounter{subsubsectionc}{0}\noindent
	{\normalsize\it\thesectionc.\thesubsectionc. #1}\par\vspace*{0.4cm}}
\renewcommand{\subsubsection}[1]
{\vspace*{0.6cm}\addtocounter{subsubsectionc}{1}
	\noindent {\normalsize\rm\thesectionc.\thesubsectionc.\thesubsubsectionc.
	#1}\par\vspace*{0.4cm}}

\newcounter{appendixc}
\newcounter{subappendixc}[appendixc]
\newcounter{subsubappendixc}[subappendixc]

\renewcommand{\appendix}[1] {\vspace*{0.6cm}
        \refstepcounter{appendixc}
        \setcounter{figure}{0}
        \setcounter{table}{0}
        \setcounter{equation}{0}
        \renewcommand{\thefigure}{\Alph{appendixc}.\arabic{figure}}
        \renewcommand{\thetable}{\Alph{appendixc}.\arabic{table}}
        \renewcommand{\theappendixc}{\Alph{appendixc}}
        \renewcommand{\theequation}{\Alph{appendixc}.\arabic{equation}}
        \noindent{\bf Appendix \theappendixc #1}\par\vspace*{0.4cm}}

\def\abstracts#1{{

\centering{\begin{minipage}{12.2truecm}\footnotesize\baselineskip=12pt\noindent
	\centerline{\footnotesize ABSTRACT}\vspace*{0.3cm}
	\parindent=0pt #1
	\end{minipage}}\par}}

\renewenvironment{thebibliography}[1]
	{\begin{list}{\arabic{enumi}.}
	{\usecounter{enumi}\setlength{\parsep}{0pt}
\setlength{\leftmargin 1.25cm}{\rightmargin 0pt}
	 \setlength{\itemsep}{0pt} \settowidth
	{\labelwidth}{#1.}\sloppy}}{\end{list}}

\newcounter{itemlistc}
\newcounter{romanlistc}
\newcounter{alphlistc}
\newcounter{arabiclistc}

\newcommand{\fcaption}[1]{
        \refstepcounter{figure}
        \setbox\@tempboxa = \hbox{\footnotesize Fig.~\thefigure. #1}
        \ifdim \wd\@tempboxa > 6in
           {\begin{center}
        \parbox{6in}{\footnotesize\baselineskip=12pt Fig.~\thefigure. #1}
            \end{center}}
        \else
             {\begin{center}
             {\footnotesize Fig.~\thefigure. #1}
              \end{center}}
        \fi}

\newcommand{\tcaption}[1]{
        \refstepcounter{table}
        \setbox\@tempboxa = \hbox{\footnotesize Table~\thetable. #1}
        \ifdim \wd\@tempboxa > 6in
           {\begin{center}
        \parbox{6in}{\footnotesize\baselineskip=12pt Table~\thetable. #1}
            \end{center}}
        \else
             {\begin{center}
             {\footnotesize Table~\thetable. #1}
              \end{center}}
        \fi}

\def\@citex[#1]#2{\if@filesw\immediate\write\@auxout
	{\string\citation{#2}}\fi
\def\@citea{}\@cite{\@for\@citeb:=#2\do
	{\@citea\def\@citea{,}\@ifundefined
	{b@\@citeb}{{\bf ?}\@warning
	{Citation `\@citeb' on page \thepage \space undefined}}
	{\csname b@\@citeb\endcsname}}}{#1}}

\newif\if@cghi
\def\cite{\@cghitrue\@ifnextchar [{\@tempswatrue
	\@citex}{\@tempswafalse\@citex[]}}
\def\citelow{\@cghifalse\@ifnextchar [{\@tempswatrue
	\@citex}{\@tempswafalse\@citex[]}}
\def\@cite#1#2{{$\null^{#1}$\if@tempswa\typeout
	{IJCGA warning: optional citation argument
	ignored: `#2'} \fi}}

 1
 1
 1

\font\ninerm=cmr9

\input{psfig}

\begin{document}

\newcommand{\st}{\scriptstyle}
\newcommand{\sst}{\scriptscriptstyle}
\newcommand{\mco}{\multicolumn}
\newcommand{\epp}{\epsilon^{\prime}}
\newcommand{\vep}{\varepsilon}
\newcommand{\ra}{\rightarrow}
\newcommand{\ppg}{\pi^+\pi^-\gamma}
\newcommand{\vp}{{\bf p}}
\newcommand{\ko}{K^0}
\newcommand{\kb}{\bar{K^0}}
\newcommand{\al}{\alpha}
\newcommand{\ab}{\bar{\alpha}}
\def\be{\begin{equation}}
\def\ee{\end{equation}}
\def\bea{\begin{eqnarray}}
\def\eea{\end{eqnarray}}
\def\CPbar{\hbox{{\rm CP}\hskip-1.80em{/}}}


\begin{flushright}
$
\begin{array}{l}
\mbox{UCD--95--3}\\[-1mm]
\mbox{January~1995}\\[2mm]
\end{array}
$
\end{flushright}
\vspace*{0.5cm}

\centerline{\normalsize\bf
Probing the standard model with hadronic $WZ$ production\footnote{Talk
given at the Beyond the Standard Model IV Conference,
Tahoe City, CA, December, 1994.}}
\baselineskip=22pt

\centerline{\footnotesize J. OHNEMUS}
\baselineskip=13pt
\centerline{\footnotesize\it Department of Physics, University of California}
\baselineskip=12pt
\centerline{\footnotesize\it Davis, CA 95616, USA}

\vspace*{0.9cm}
\abstracts{
The cross section for producing $WZ$ pairs at hadron colliders is
calculated to order $\alpha_s$ for general $C$ and $P$ conserving $WWZ$
couplings. The effects of the next-to-leading-order corrections on the
derived sensitivity limits for anomalous $WWZ$ couplings are
discussed.  The prospects for observing the approximate amplitude zero, which
is present in the standard model $WZ$ helicity amplitudes, are also discussed.}

\normalsize\baselineskip=15pt
\setcounter{footnote}{0}
\renewcommand{\thefootnote}{\alph{footnote}}

\section{Introduction}

The production of $WZ$ pairs at hadron colliders provides an
excellent opportunity to study the $WWZ$
vertex\cite{FIRSTWZ,WWZ,UJH}.  In addition,
the reaction $p\,p\hskip-7pt\hbox{$^{^{(\!-\!)}}$} \rightarrow
W^{\pm}Z$ is of interest due to the presence of an approximate zero in the
amplitude of the
parton level subprocess $q_1\bar q_2\rightarrow W^\pm
Z$\cite{UJH} in the standard model,
which is similar in nature to the well-known
radiation zero in the reaction
$p\,p\hskip-7pt\hbox{$^{^{(\!-\!)}}$} \rightarrow
W^{\pm}\gamma$\cite{RAZ}.
Previous studies on probing the $WWZ$ vertex via hadronic $WZ$
production have been based on leading-order
calculations\cite{WWZ}.
This report summarizes the results of a comprehensive study\cite{WZPRD}
of hadronic
$WZ$ production based on an
${\cal O}(\alpha_s)$ calculation of the reaction
$p\,p\hskip-7pt\hbox{$^{^{(\!-\!)}}$} \rightarrow W^{\pm}Z + X
\rightarrow \ell_1^\pm \nu_1 \ell_2^+ \ell_2^- + X$ for general,
$C$ and $P$ conserving, $WWZ$ couplings.

\section{Anomalous Couplings}

In the standard model (SM), the $WWZ$ vertex is uniquely determined
by the $\hbox{\rm SU(2)} \bigotimes \hbox{\rm U(1)}$ gauge
structure of the electroweak sector, thus a measurement of the $WWZ$
vertex provides a stringent test of the SM.
The most general $WWZ$ vertex,
which is Lorentz, $C$, and $P$ invariant,
contains three free parameters, $g_1, \kappa$, and $\lambda$,
and is described by the effective Lagrangian\cite{LAGRANGIAN}
%
%
\[
{\cal L}_{WWZ} = -i\, e\, {\rm cot} \theta_{\rm W}
\Bigl[ g_1 \bigl( W_{\mu\nu}^{\dagger} W^{\mu} Z^{\nu}
                  -W_{\mu}^{\dagger} Z_{\nu} W^{\mu\nu} \bigr)
+ \kappa W_{\mu}^{\dagger} W_{\nu} Z^{\mu\nu}
+ {\lambda \over M_W^2} W_{\lambda \mu}^{\dagger} W^{\mu}_{\nu}
Z^{\nu\lambda} \Bigr] .
\]
%
%
At tree level in the SM, $g_1=1$, $\kappa = 1$, and $\lambda = 0$.

The $Z$ boson transverse momentum
spectrum is very sensitive to anomalous $WWZ$ couplings.
At the Tevatron, the
${\cal O}(\alpha_s)$ QCD corrections are modest and
sensitivities are only slightly affected by the QCD corrections.
At the
LHC, on the other hand, the inclusive ${\cal O}(\alpha_s)$ QCD
corrections in the SM are
very large at high $p_T^{}(Z)$, and have a
non-negligible influence on the sensitivity bounds which can be achieved
for anomalous $WWZ$ couplings; compare Figs.~1a and 1b.
The large QCD corrections are caused by
the combined effects of destructive interference in the Born subprocess,
a log-squared enhancement factor in the $q_1 g \rightarrow W Z q_2$
partonic cross section at high
transverse momentum\cite{FRIX}, and the large quark-gluon luminosity
at supercollider energies.
The size of the QCD corrections at high
$p_T^{}(Z)$ can be significantly reduced, and a
significant fraction of the sensitivity lost at the LHC energy can be
regained, if a jet veto
is imposed, {\it i.e.}, if the $WZ+0$~jet exclusive channel is considered;
see Fig.~1c.

\begin{figure}
\begin{center}
\leavevmode
\psfig{figure=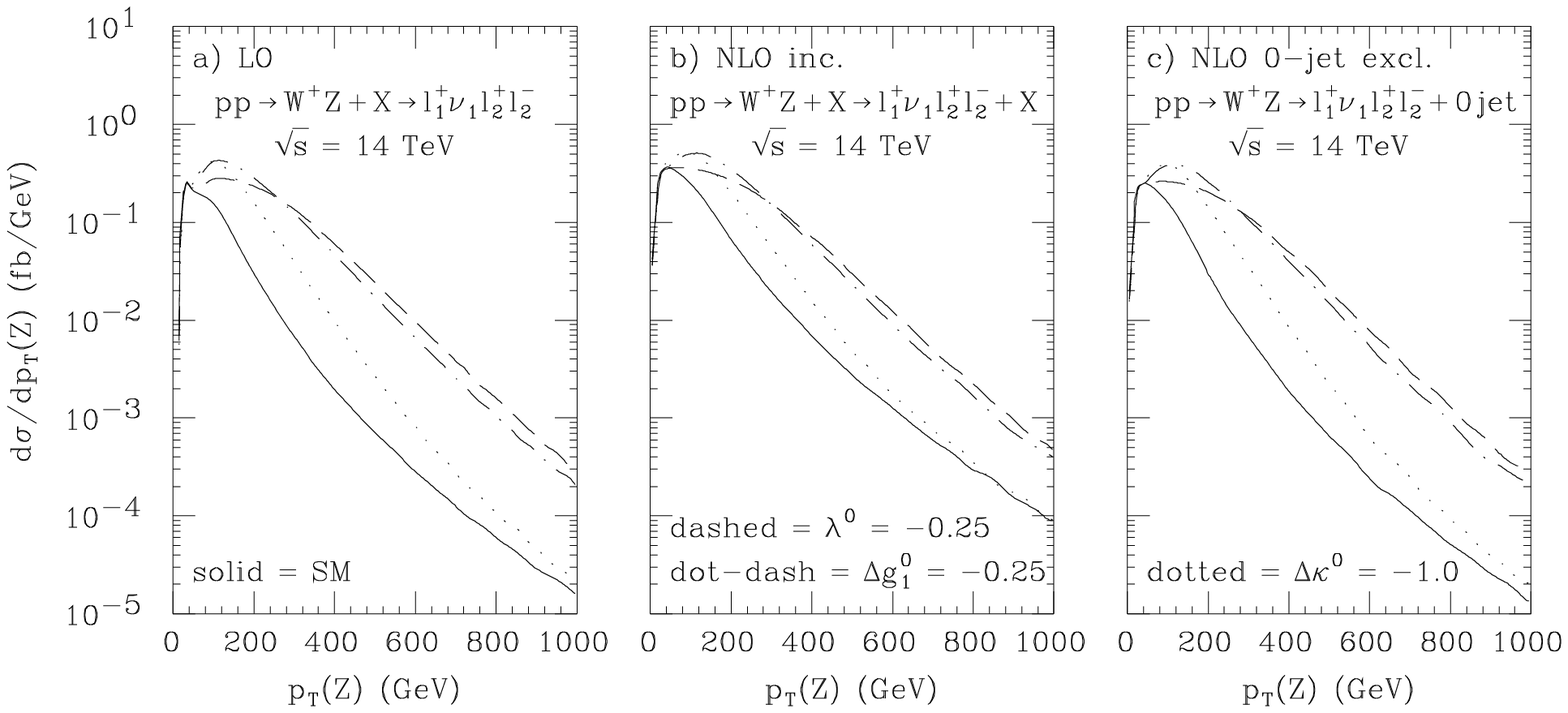,width=6.0in,clip=}
\fcaption{Transverse momentum distribution of the $Z$ boson
for the standard model and three values of anomalous couplings.
Parts a), b), and c) are the results for the LO, NLO inclusive,
and the NLO 0-jet exclusive cross sections, respectively.}
\end{center}
\label{fig:one}
\end{figure}

\section{Approximate Amplitude Zero}

Recently, it has been shown that the SM amplitude for $q_1\bar q_2
\rightarrow W^\pm Z$ at the Born level exhibits an approximate zero
at high energies, $\hat s\gg M_Z^2$, located at\cite{UJH}
%
%
\[
\cos\Theta^* \approx \pm {1\over 3}\tan^2\theta_{\rm W}
\approx \pm 0.1,
\]
%
%
where $\Theta^*$ is the scattering
angle of the $Z$ boson relative to the quark direction in the $WZ$
center of mass frame. The approximate zero is the combined result of
an exact zero in the dominant helicity amplitudes ${\cal M}(\pm,\mp)$ and
strong gauge cancellations in the remaining amplitudes.

The approximate amplitude zero in $WZ$
production causes a slight dip in the
rapidity difference distribution,
$\Delta y(Z,\ell_1)=y(Z)-y(\ell_1)$, where $\ell_1$ is the charged lepton
from the decaying $W$ boson; see Fig.~2a.
At the Tevatron
energy, order $\alpha_s$ QCD corrections have a negligible influence the
shape of the $\Delta y(Z,\ell_1)$ distribution. At the LHC, however,
${\cal O}(\alpha_s)$
QCD effects completely obscure the dip, unless a jet veto is
imposed.

Cross section ratios can also be used to search for
experimental consequences of the approximate amplitude zero.
The ratio of $ZZ$ to $WZ$ cross sections as a function of the minimum
$Z$ boson transverse momentum, $p_T^{\rm min}$, increases with $p_T^{\rm
min}$ for values larger than 100~GeV; see Fig.~2b.
This increase in the cross section ratio
is a direct consequence of the approximate zero.
QCD corrections
have a significant impact on the $ZZ$ to $WZ$ cross section ratio at
the LHC unless a jet veto is imposed.

\begin{figure}
\begin{center}
\leavevmode
\psfig{figure=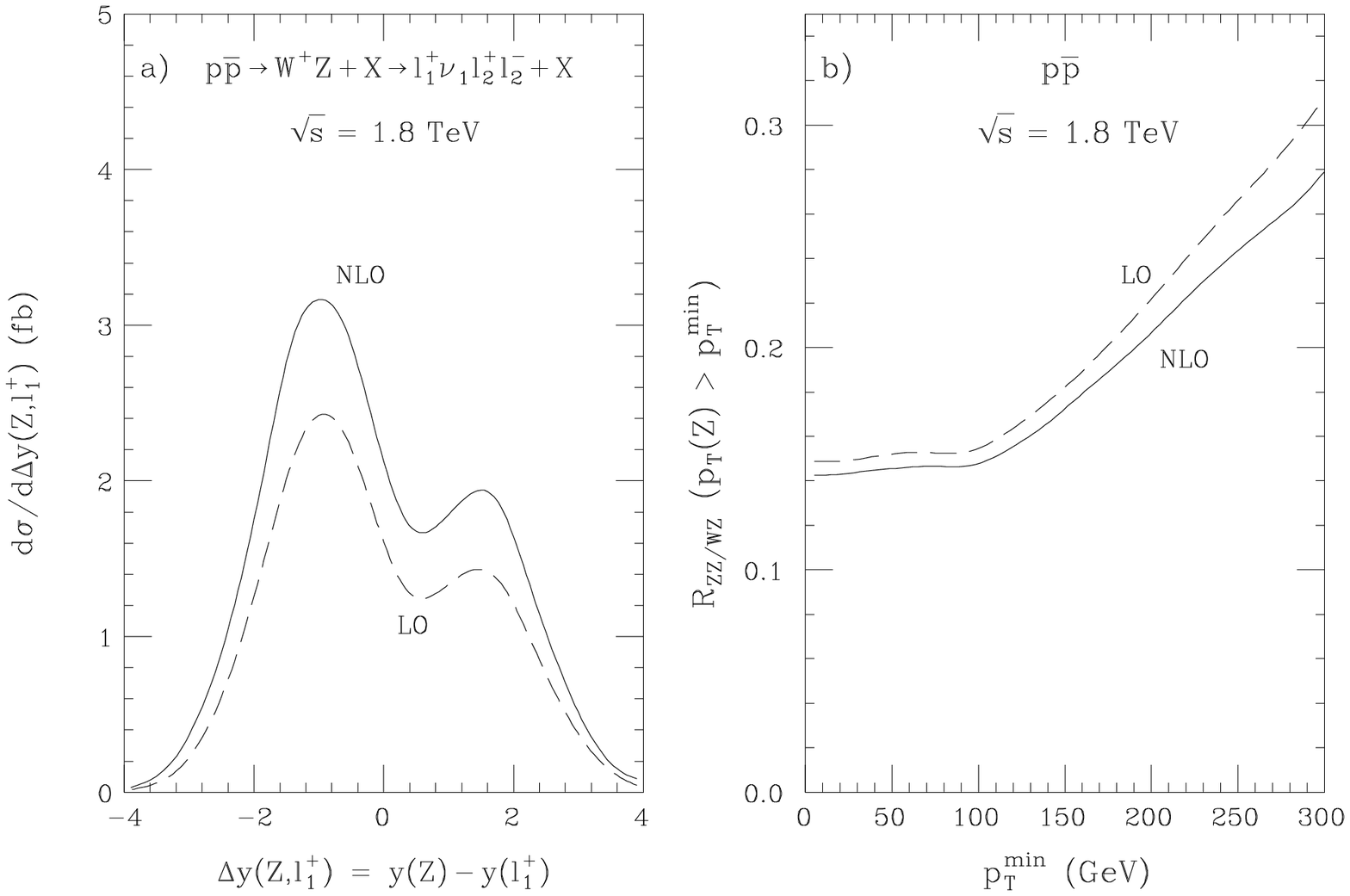,width=3.5in,clip=}
\fcaption{a) Distribution of the rapidity difference $y(Z) - y(\ell_1^+)$.
b) The $ZZ$ to $WZ$ cross section ratio as a function of
the minimum transverse momentum of the $Z$ boson.}
\end{center}
\label{fig:two}
\end{figure}

The $\Delta y(Z,\ell_1)$ distribution and the $ZZ$ to $W^\pm Z$
cross section ratio are useful tools in
searching for
the approximate amplitude zero in $WZ$ production. However, for
the integrated luminosities envisioned, it will not be easy
to conclusively establish the approximate amplitude zero in $WZ$
production at the Tevatron or the LHC.

\section{Acknowledgements}

Collaborations with U.~Baur and T.~Han on this work are gratefully
acknowledged.
This work has been supported in part by Department of Energy
grant No.~DE-FG03-91ER40674.


\normalsize

\section{References}

\begin{thebibliography}{9}
%
\bibitem{FIRSTWZ}
R.~Brown, K.~Mikaelian, and D.~Sahdev,
{\it Phys. Rev.} {\bf D20} (1979) 1164.
%
\bibitem{WWZ}
D.~Zeppenfeld and S.~Willenbrock,
{\it Phys. Rev.} {\bf D37} (1988) 1775;
M.~Kuroda, J.~Maalampi, K.~Schwarzer, and D.~Schildknecht,
{\it Nucl. Phys.} {\bf B284} (1987) 271;
S.-C.~Lee and W.-C.~Su, {\it Phys. Lett.} {\bf B212} (1988) 113;
K.~Hagiwara, J.~Woodside, and D.~Zeppenfeld,
{\it Phys. Rev.} {\bf D41} (1990) 2113;
H.~Kuijf {\it et~al.}, {\it Proceedings of the ECFA
Workshop on LHC Physics}, Aachen, FRG, 1990, Vol.~II, p.~91. 
%
\bibitem{UJH}
U.~Baur, T.~Han, and J.~Ohnemus, {\it Phys. Rev. Lett.} {\bf 72} (1994) 3941.
%
\bibitem{RAZ}
K.~Mikaelian, M.~Samuel, and D.~Sahdev,
{\it Phys. Rev. Lett.} {\bf 43} (1979) 746. 	
%
\bibitem{WZPRD}
U.~Baur, T.~Han, and J.~Ohnemus, to appear in {\it Phys. Rev.} {\bf D}.
%
\bibitem{LAGRANGIAN}
K.~Hagiwara, R.~D.~Peccei, D.~Zeppenfeld, and K.~Hikasa,
{\it Nucl. Phys.} {\bf B282} (1987) 253;
K.~Gaemers and G.~Gounaris,
{\it Z. Phys.} {\bf C1} (1979) 259. 		
%
\bibitem{FRIX}
S.~Frixione, P.~Nason, and G.~Ridolfi,
{\it Nucl. Phys.} {\bf B383} (1992) 3.  
%
\end{thebibliography}
\end{document}